\documentclass[conference]{IEEEtran}
\IEEEoverridecommandlockouts
\usepackage{cite}
\usepackage{amsmath,amssymb,amsfonts}
\usepackage{graphicx}
\usepackage{textcomp}
\usepackage{xcolor}
\usepackage{booktabs}
\usepackage{url}
\def\BibTeX{{\rm B\kern-.05em{\sc i\kern-.025em b}\kern-.08em
    T\kern-.1667em\lower.7ex\hbox{E}\kern-.125emX}}

\title{A Vector Metadata-Residency Subsystem for Grouped-Query Attention on FPGA Accelerators}

\author{\IEEEauthorblockN{Venkata Sai Rajesh Malladi}
\IEEEauthorblockA{\textit{Digital Design Engineer} \\
\textit{Texas Instruments} \\
Bengaluru, India}}

\begin{document}
\maketitle

\begin{abstract}
Long-context inference systems increasingly require hardware support for deciding which token or KV-cache metadata should remain resident near the accelerator. This paper extends a scalar non-stalling metadata-residency tile into a scalable vector subsystem for per-layer, per-head, and grouped-query-attention (GQA) deployment. The scalar tile performs one metadata decision per cycle using dynamic threshold admission, structural hashing, a dual-bank BRAM table, a small CAM stash, bounded score-conditioned eviction, and non-blocking bypass/drop paths. A new 1024-bit vector wrapper instantiates sixteen independent scalar tiles behind sixteen 64-bit lanes, allowing either eight waves for 128 query heads or one wave for sixteen KV groups under an 8:1 GQA mapping when upstream logic emits per-KV-group metadata. Out-of-context synthesis of the vector1024 subsystem on a Xilinx Kintex UltraScale xcku13p device meets a 200 MHz target with +0.255 ns WNS, using 28698 LUTs, 25955 registers, 48 RAMB36E2 blocks, and zero DSPs. Functional simulation passes for the 1024-bit/16-lane wrapper. Trace-level evaluation on Qwen2.5 and DeepSeek attention metadata shows full maximum-score top-K recall by $K=16$ in the evaluated grouped schedule; a separate exhaustive DeepSeek-R1 1.5B archive sweep over 28 layers, 12 heads per layer, and 8192 events per head shows high mean maximum-score recall across all 336 traces. The result is metadata-residency and scheduling evidence, not an end-to-end LLM quality or GPU speedup claim.
\end{abstract}

\section{Introduction}
Modern transformer inference pipelines process many streams of token metadata: token positions, attention scores, cache slots, layer identifiers, head identifiers, and routing flags. In long-context decoding, these metadata streams influence which KV-cache entries are accessed, retained, spilled, or ignored. Keeping all metadata and all payloads resident on chip is not practical, but pushing every low-value metadata item to external memory also wastes bandwidth. The design problem addressed here is therefore \emph{line-rate metadata residency}: every cycle, hardware must decide whether incoming metadata should update local state, allocate a new resident entry, evict a weaker resident, bypass to an external path, or drop locally.

This work builds on a scalar metadata-residency tile and evaluates a total vector subsystem. The scalar tile stores compact token metadata in FPGA BRAM and makes bounded local decisions without recursive cuckoo insertion or a global sorter. The vector subsystem replicates this tile across lanes so multiple heads or KV groups can be processed in parallel. The resulting design is intended as an FPGA-side companion to attention engines, DMA engines, and memory controllers: it does not compute attention, but it maintains score-aware metadata state at fabric rate.

The paper makes five contributions:
\begin{itemize}
    \item A total-system organization that scales the scalar metadata-residency tile across layers, heads, and vector lanes.
    \item A 1024-bit, sixteen-lane RTL implementation in which each lane consumes a 64-bit metadata word and instantiates an independent tile.
    \item A GQA scheduling evaluation showing that 128 query heads map to eight vector1024 waves in raw query-head mode, while sixteen KV groups under 8:1 GQA fit in one vector1024 wave if upstream aggregation is available.
    \item A full DeepSeek-R1-Distill-Qwen-1.5B (``DeepSeek-R1 1.5B'') trace sweep over all detected layer/head CSVs in the provided archive, with CSV summaries and graphs for recall, precision, sink retention, drop rate, and layer/head behavior.
    \item Post-synthesis FPGA evidence: the vector1024 subsystem meets a 5 ns OOC synthesis target with +0.255 ns WNS and no DSP usage.
\end{itemize}

\section{Related Work}
\subsection{Top-K and Frequent-Item Algorithms}
Misra and Gries gave early algorithms for finding repeated elements~\cite{misra1982}. Later stream methods address frequent-item sketching~\cite{charikar2002}, approximate frequency counts~\cite{manku2002}, frequent and top-$k$ elements using Space-Saving~\cite{metwally2005spacesaving}, and general stream summaries such as the Count-Min sketch~\cite{cormode2005countmin}. HeavyKeeper targets top-$k$ elephant flows~\cite{gong2018heavykeeper}. These works motivate bounded-state selection. The tile in this paper ranks metadata by attention score rather than occurrence frequency, and it does not compute an exact top-$K$.

\subsection{Hardware Selection and Hashing}
Teubner \emph{et al.} study frequent-item computation on FPGAs~\cite{teubner2011}, while HashPipe detects heavy hitters in a programmable-switch data plane using pipelined hash tables~\cite{sivaraman2017}. The two-choice placement used here is related to cuckoo hashing~\cite{pagh2004cuckoo}, and its overflow storage is related to cuckoo hashing with a stash~\cite{kirsch2008}; the tile does not perform recursive cuckoo insertion.

\subsection{Token and KV-Cache Selection}
SpAtten applies cascade token and head pruning in a sparse attention architecture~\cite{wang2021spatten}. H$_2$O~\cite{zhang2023h2o}, StreamingLLM~\cite{xiao2024streamingllm}, SnapKV~\cite{li2024snapkv}, Quest~\cite{tang2024quest}, and Ada-KV~\cite{feng2025adakv} study which cached information to keep or access during efficient inference. This paper examines a narrower question: score-conditioned metadata residency in an FPGA subsystem. No head-to-head superiority over these methods is claimed.

\section{System Context}
The subsystem is designed for a streaming accelerator fabric rather than a monolithic CPU cache. Each metadata event contains an identity, score, pointer, and optional routing information. The identity is used for table update and collision handling, the score drives threshold admission and replacement, and the pointer refers to the external payload or KV-cache slot. Full KV vectors remain outside this block.

\subsection{Attention Parallelism}
Transformer heads inside a layer are independent during attention score generation and value aggregation. After all heads in a layer complete, the outputs are concatenated and transformed through the output projection, residual path, and subsequent MLP or MoE block. Layers are therefore sequential at the model level, while heads or KV groups within a layer are the natural unit of metadata parallelism. The subsystem should be interpreted as per-layer head or KV-group metadata filtering, not as a pipeline that sends a head directly through all model layers.

\subsection{Query-Head and KV-Group Modes}
The 1024-bit vector wrapper exposes sixteen 64-bit lanes. In raw query-head mode, each lane maps to one query head. A 128-head layer therefore requires
\begin{equation}
\left\lceil \frac{128}{16} \right\rceil = 8
\end{equation}
vector waves per layer. In GQA mode~\cite{ainslie2023gqa}, multiple query heads share one KV head or KV group. With an 8:1 query-to-KV mapping, 128 query heads map to sixteen KV groups, and the same vector1024 wrapper covers all groups in one wave:
\begin{equation}
\left\lceil \frac{128/8}{16} \right\rceil = 1.
\end{equation}
The traced distilled models use the Qwen2 architecture. DeepSeek-R1-Distill-Qwen-1.5B has 28 layers, 12 query heads, and 2 key-value heads (6:1), and DeepSeek-R1-Distill-Qwen-7B has 28 query heads and 4 key-value heads (7:1).\footnote{Model \texttt{config.json} files at \url{https://huggingface.co/deepseek-ai}.} The 8:1 mapping above is therefore a hardware scheduling projection, not the native ratio of these models. This one-wave claim is conditional. It requires upstream attention logic to produce or aggregate one metadata stream per KV group before the metadata-residency subsystem. It should not be read as a claim that the current RTL internally combines eight raw query-head streams into one KV-group stream.

\section{Scalar Tile}
The scalar tile accepts one metadata record per cycle and uses a deterministic bounded policy. It is not an exact top-K sorter. Instead, it preserves a score-aware resident set using threshold admission and two-choice table placement.

\subsection{Metadata Format}
The scalar tile currently stores a 72-bit internal slot:
\begin{equation}
\{token\_id[31:0], score[15:0], ptr[23:0]\}.
\end{equation}
The pointer represents a sequence position, cache slot, or external memory reference. The scalar RTL computes a structural identity
\begin{equation}
inst\_key = token\_id \oplus ptr.
\end{equation}
This prevents repeated vocabulary IDs at different sequence positions from collapsing into one resident entry. In vector mode, the wrapper ties \texttt{token\_id} to zero and uses \texttt{ptr} as the effective identity, so \(inst\_key=ptr\). A future storage optimization can remove the unused token ID and store a 48-bit slot containing pointer, score, and flags, but that narrower slot is not part of the measured RTL reported here.

\subsection{Decision Policy}
The scalar tile performs the following steps each cycle. First, a Top-K-inspired gate converts the signed score into a monotonic key and updates an IIR threshold. Second, two seeded hashes generate candidate addresses in two BRAM banks. Third, the victim selector resolves hits, empty slots, and collisions. A hit updates the resident slot, an empty candidate admits the new slot, and a collision can evict a lower-score resident into a FIFO. A four-entry CAM stash absorbs small collision bursts. If the stash has a hit it updates; if an empty stash entry exists it inserts; if the stash is full the incoming item bypasses. This simplified stash policy was selected because a previous min-score stash replacement path missed timing.

\section{Vector and Total-System RTL}
\begin{figure*}[t]
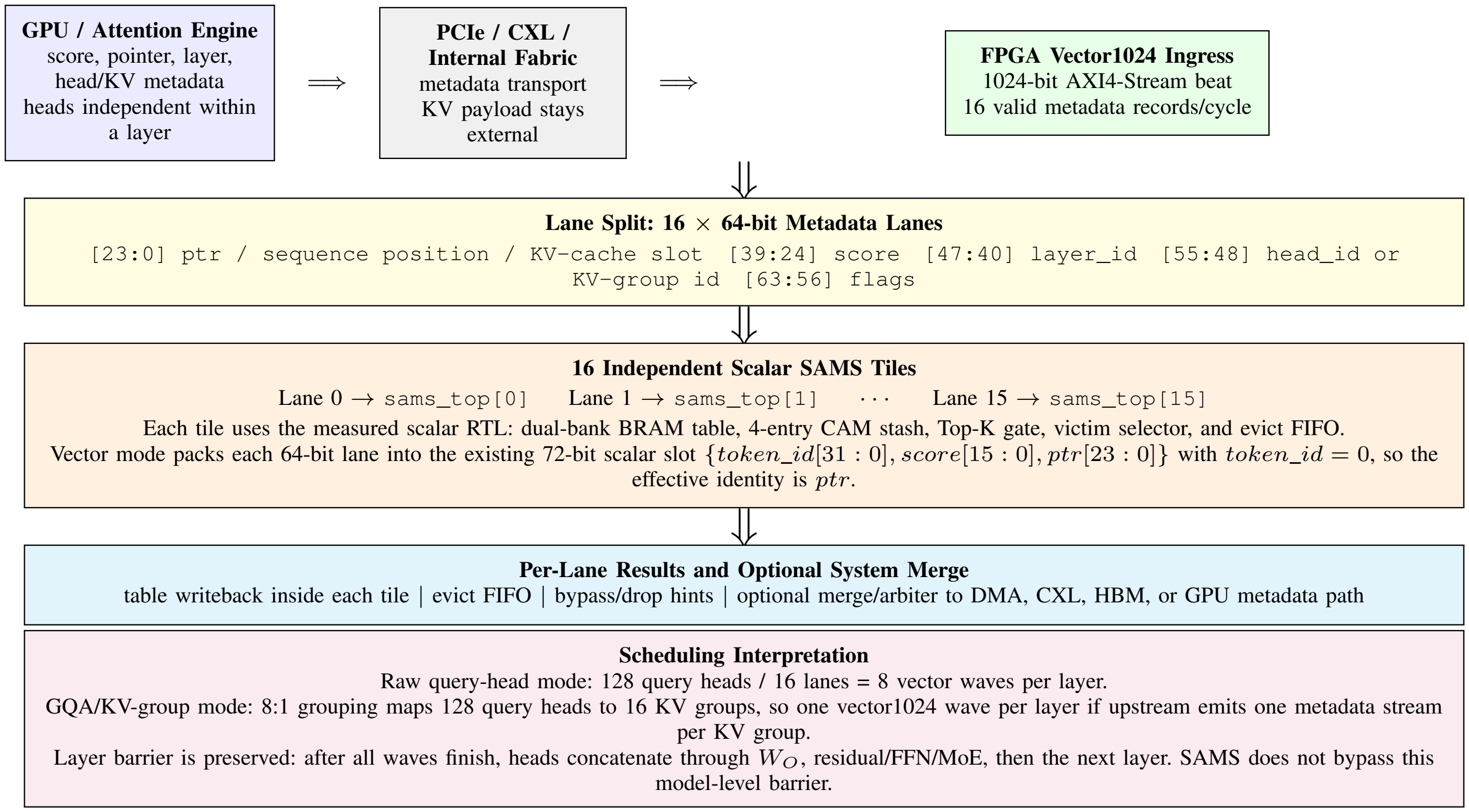

\centering
\scriptsize
\setlength{\fboxsep}{5pt}
\resizebox{0.98\textwidth}{!}{%
\begin{tabular}{@{}c@{\hspace{0.35cm}}c@{\hspace{0.35cm}}c@{\hspace{0.35cm}}c@{\hspace{0.35cm}}c@{}}
\fcolorbox{black}{blue!8}{\parbox{2.6cm}{\centering\textbf{GPU / Attention Engine}\\score, pointer, layer, head/KV metadata\\heads independent within a layer}}
& $\Longrightarrow$
& \fcolorbox{black}{gray!12}{\parbox{2.35cm}{\centering\textbf{PCIe / CXL / Internal Fabric}\\metadata transport\\KV payload stays external}}
& $\Longrightarrow$
& \fcolorbox{black}{green!10}{\parbox{3.2cm}{\centering\textbf{FPGA Vector1024 Ingress}\\1024-bit AXI4-Stream beat\\16 valid metadata records/cycle}}
\\[0.45cm]
\multicolumn{5}{c}{\Large$\Downarrow$}\\[-0.05cm]
\multicolumn{5}{c}{%
\fcolorbox{black}{yellow!14}{\parbox{15.5cm}{\centering
\textbf{Lane Split: 16 $\times$ 64-bit Metadata Lanes}\\[0.05cm]
\texttt{[23:0] ptr / sequence position / KV-cache slot}\quad
\texttt{[39:24] score}\quad
\texttt{[47:40] layer\_id}\quad
\texttt{[55:48] head\_id or KV-group id}\quad
\texttt{[63:56] flags}}}}
\\[0.35cm]
\multicolumn{5}{c}{\Large$\Downarrow$}\\[-0.05cm]
\multicolumn{5}{c}{%
\fcolorbox{black}{orange!12}{\parbox{15.5cm}{\centering
\textbf{16 Independent Scalar SAMS Tiles}\\[0.05cm]
\begin{tabular}{cccc}
Lane 0 $\rightarrow$ \texttt{sams\_top[0]} & Lane 1 $\rightarrow$ \texttt{sams\_top[1]} & $\cdots$ & Lane 15 $\rightarrow$ \texttt{sams\_top[15]}\\
\end{tabular}\\[0.05cm]
Each tile uses the measured scalar RTL: dual-bank BRAM table, 4-entry CAM stash, Top-K gate, victim selector, and evict FIFO.\\
Vector mode packs each 64-bit lane into the existing 72-bit scalar slot \(\{token\_id[31:0], score[15:0], ptr[23:0]\}\) with \(token\_id=0\), so the effective identity is \(ptr\).}}}
\\[0.35cm]
\multicolumn{5}{c}{\Large$\Downarrow$}\\[-0.05cm]
\multicolumn{5}{c}{%
\fcolorbox{black}{cyan!10}{\parbox{15.5cm}{\centering
\textbf{Per-Lane Results and Optional System Merge}\\
table writeback inside each tile $\mid$ evict FIFO $\mid$ bypass/drop hints $\mid$ optional merge/arbiter to DMA, CXL, HBM, or GPU metadata path}}}
\\[0.35cm]
\multicolumn{5}{c}{%
\fcolorbox{black}{purple!8}{\parbox{15.5cm}{\centering
\textbf{Scheduling Interpretation}\\
Raw query-head mode: 128 query heads / 16 lanes = 8 vector waves per layer.\\
GQA/KV-group mode: 8:1 grouping maps 128 query heads to 16 KV groups, so one vector1024 wave per layer if upstream emits one metadata stream per KV group.\\
Layer barrier is preserved: after all waves finish, heads concatenate through \(W_O\), residual/FFN/MoE, then the next layer. SAMS does not bypass this model-level barrier.}}}
\end{tabular}}
\caption{Total-system microarchitecture for the vector metadata-residency subsystem. A 1024-bit AXI4-Stream beat is split into sixteen 64-bit metadata lanes, each feeding an independent instance of the measured scalar \texttt{sams\_top} tile. The diagram distinguishes raw query-head scheduling from KV-group scheduling and shows that downstream merging is an integration option outside the scalar tile.}
\label{fig:system_microarch}
\end{figure*}

Fig.~\ref{fig:system_microarch} shows the complete vector metadata-residency subsystem around the replicated scalar tiles. The attention engine, GPU, or a near-memory accelerator emits metadata events into an FPGA-side fabric through PCIe/CXL or an internal accelerator interconnect. The vector wrapper consumes a 1024-bit AXI4-Stream beat, slices it into sixteen independent 64-bit metadata lanes, and statically maps each lane to one \texttt{sams\_top} tile. Each tile performs its own threshold admission, two-choice table lookup, stash handling, and bounded eviction without cross-lane dependence. Outputs remain per-lane unless a downstream merge, DMA, CXL, HBM, or GPU return path is instantiated by the surrounding system.

\begin{figure*}[t]
\centering
\includegraphics[width=0.98\textwidth]{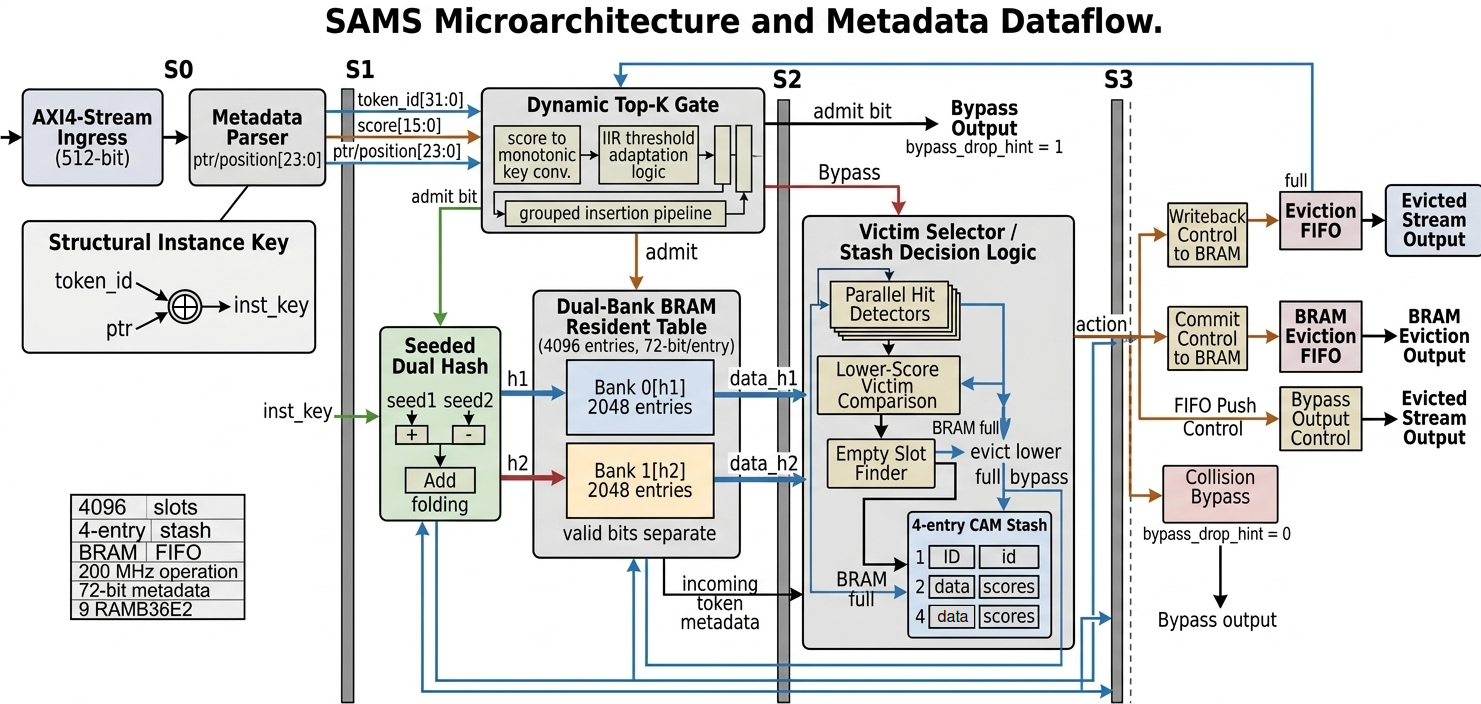}
\caption{Scalar metadata-residency tile used as the vector subsystem building block. The vector1024 wrapper instantiates sixteen independent copies of this tile, one per 64-bit lane.}
\label{fig:microarch}
\end{figure*}

Fig.~\ref{fig:microarch} shows the scalar microarchitecture that is replicated inside each lane of Fig.~\ref{fig:system_microarch}.

\subsection{1024-bit Lane Format}
The vector1024 top splits a 1024-bit AXI4-Stream beat into sixteen 64-bit lanes. Each lane has the following format:
\begin{align*}
\texttt{[23:0]}   & : \text{pointer, sequence position, or KV-cache slot},\\
\texttt{[39:24]}  & : \text{score},\\
\texttt{[47:40]}  & : \text{layer identifier},\\
\texttt{[55:48]}  & : \text{head or KV-group identifier},\\
\texttt{[63:56]}  & : \text{flags}.
\end{align*}
Each lane feeds one scalar tile. The current RTL forwards layer, head, and flags as reserved metadata fields while the scalar tile stores the 72-bit \(token\_id,score,ptr\) record with \(token\_id=0\) for vector-mode identity.

\subsection{Scalable System Wrapper}
The scalable system top provides an alternative per-layer/per-head dispatch structure. It accepts one global event per cycle and routes it to a selected layer/head tile. It supports per-tile threshold seeds, per-tile hash seeds, and optional pinned sink entries. Pinned sink entries are intended for attention-sink policies such as preserving BOS or similar high-importance positions. This wrapper is useful for control-rich deployments, while the vector1024 wrapper is the high-throughput configuration evaluated for 1024-bit data movement.

\section{Verification and Synthesis}
The 1024-bit wrapper was functionally verified with a smoke test that drives sixteen lanes in a 1024-bit AXI beat. The simulation reports:
\begin{quote}\footnotesize\raggedright
\texttt{[PASS] sams\_vector1024\_top 1024-bit / 16-lane smoke test completed}
\end{quote}

Out-of-context synthesis was run using Vivado 2016.2 on xcku13p-ffve900-1-e-EVAL with a 5 ns target. Table~\ref{tab:vector_hw} summarizes the measured vector1024 result. The measured scalar tile is also shown for reference. Both results are OOC synthesis snapshots of the simplified-stash configuration, not full place-and-route timing for the vector subsystem.

\begin{table}[t]
\centering
\caption{Measured OOC synthesis results at 200 MHz.}
\label{tab:vector_hw}
\begin{tabular}{lrr}
\toprule
Metric & Scalar tile & Vector1024 subsystem \\
\midrule
AXI width & 512 bit & 1024 bit \\
Lanes & 1 & 16 \\
CLB LUTs & 6105 & 28698 \\
CLB registers & 5295 & 25955 \\
RAMB36E2 & 9 & 48 \\
DSPs & 0 & 0 \\
WNS at 5 ns & +0.106 ns & +0.255 ns \\
TNS & 0.000 ns & 0.000 ns \\
\bottomrule
\end{tabular}
\end{table}

The vector1024 critical path starts at a lane-local resident-table BRAM clock and ends at a lane-local commit write-address register. The reported data path delay is 4.594 ns with eight logic levels and +0.255 ns slack. This confirms that the replicated lane structure preserves a 200 MHz OOC synthesis target without DSPs.

\section{Evaluation Methodology}
The evaluation uses the existing policy-accurate Python model of the RTL state transitions. The model includes score-key conversion, threshold adaptation, seeded hashing, dual-bank resident tables, score-conditioned eviction, FIFO pressure behavior, and simplified stash insertion/update. For vector-mode evaluation, token identity is carried by the pointer field, matching the vector wrapper where \texttt{token\_id=0}.

We evaluate four classes of evidence. First, synthetic Zipfian and baseline comparisons characterize the scalar policy. Second, empirical attention traces from Qwen2.5~\cite{qwen2024} and DeepSeek-R1~\cite{deepseekr1} 1.5B/7B model runs measure top-K recall, precision, sink recall, and drop rate. Third, the new GQA grouping evaluation projects how many vector1024 waves are required under raw query-head and 8:1 KV-group scheduling. Fourth, a full DeepSeek-R1 1.5B archive sweep evaluates every detected layer/head trace in the trace archive using the vector-mode identity convention.

\subsection{GQA Evaluation Script}
The evaluation script is \texttt{eval/run\_gqa\_group\_eval.py}. It reads user-supplied trace CSVs, for example with \texttt{--globs "traces\_1p5b/*.csv,traces\_7b/*.csv"}, evaluates each observed layer/head stream with the vector-mode identity convention, then groups reported rows by model, size, layer, KV group, and \(K\). The generated CSV is \texttt{eval/gqa\_group\_eval.csv}. The script reports both scheduling constants and accuracy metrics.

Because the available trace set is representative rather than exhaustive, most observed KV groups contain one traced head. The scheduling result is therefore a system projection calibrated by real per-head metadata traces, not a complete all-head LLM run. The traced DeepSeek-R1 models are distilled Qwen2 variants. They use GQA and not the multi-head latent attention (MLA) introduced in DeepSeek-V2~\cite{deepseekv2}; the 8:1 schedule does not match their native grouping ratios.

\section{Evaluation Results}
Table~\ref{tab:gqa_eval} summarizes the GQA grouping run. The key scheduling result is independent of trace distribution: with 128 query heads and sixteen vector lanes, query-head mode requires eight waves per layer, while 8:1 GQA mode produces sixteen KV groups and needs one vector1024 wave per layer when upstream aggregation emits group metadata.

Table~\ref{tab:key_findings} summarizes the main findings that connect the RTL, scheduling model, and trace evaluation. The hardware evidence demonstrates that lane replication is practical at the target frequency; the trace evidence shows that the policy keeps the high-score metadata set while discarding a substantial fraction of low-score events; and the scheduling projection quantifies the value of presenting grouped metadata to the vector interface.

\begin{table*}[t]
\centering
\caption{Key findings from the vector metadata-residency study.}
\label{tab:key_findings}
\begin{tabular}{p{0.12\textwidth}p{0.30\textwidth}p{0.50\textwidth}}
\toprule
Finding & Evidence & Research implication \\
\midrule
F1: Replication closes timing & Vector1024 OOC synthesis meets 200 MHz with +0.255 ns WNS, 48 RAMB36E2 blocks, and zero DSPs. & Independent lane replication is a viable FPGA organization for metadata-residency decisions rather than a timing-prohibitive global sorter. \\
F2: GQA reduces waves & 128 raw query heads require 8 waves on 16 lanes; sixteen 8:1 KV groups require 1 wave if upstream emits group metadata. & The block exposes the system benefit of grouped attention by converting architectural head sharing into fewer metadata-residency waves. \\
F3: Full archive recall remains high & DeepSeek-R1 1.5B sweep covers 28 layers, 12 heads/layer, 8192 events/head; mean maximum-score recall is 0.996 at $K=16$. & The bounded RTL policy preserves the dominant score-ranked metadata across a complete layer/head trace grid. \\
F4: Filtering remains significant & Full-archive drop rate is 0.393 at $K=16$ and 0.340 at $K=64$. & The threshold path continues to reduce low-score metadata pressure even when the resident target is enlarged. \\
F5: Claims are bounded & Results are OOC synthesis and trace-level metadata evaluation; the 8:1 mapping is a projection, not the traced models' native ratio. & The contribution is a hardware substrate and scheduling/evaluation path, not an end-to-end LLM quality or latency claim. \\
\bottomrule
\end{tabular}
\end{table*}

\begin{table}[t]
\centering
\caption{GQA/KV-group evaluation over available Qwen2.5 and DeepSeek trace CSVs.}
\label{tab:gqa_eval}
\begin{tabular}{rrrrrr}
\toprule
$K$ & Groups & Mean recall & Min recall & Sink recall & Drop rate \\
\midrule
4  & 17 & 0.985 & 0.750 & 1.000 & 0.666 \\
8  & 17 & 0.985 & 0.750 & 1.000 & 0.665 \\
16 & 17 & 1.000 & 1.000 & 1.000 & 0.664 \\
32 & 17 & 1.000 & 1.000 & 1.000 & 0.659 \\
64 & 17 & 1.000 & 1.000 & 1.000 & 0.608 \\
\bottomrule
\end{tabular}
\end{table}

The trace results show that the evaluated metadata streams reach full maximum-score top-K recall by \(K=16\). Sink recall remains 1.000 in the grouped run. Drop rate remains high because the threshold gate rejects low-attention metadata events; this is the intended bandwidth-saving mode. The result should be read as evidence that the metadata policy retains high-scoring candidates under the evaluated traces while reducing resident updates and downstream metadata pressure.

\begin{figure*}[t]
\centering
\parbox[t]{0.49\textwidth}{\centering\includegraphics[width=\linewidth]{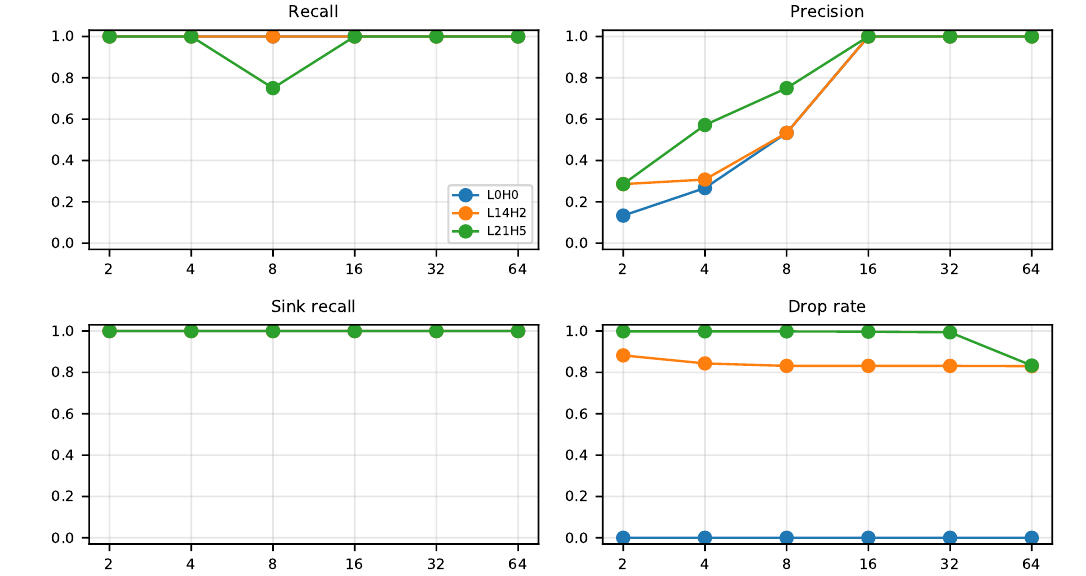}\\\footnotesize (a) Qwen2.5-1.5B}
\parbox[t]{0.49\textwidth}{\centering\includegraphics[width=\linewidth]{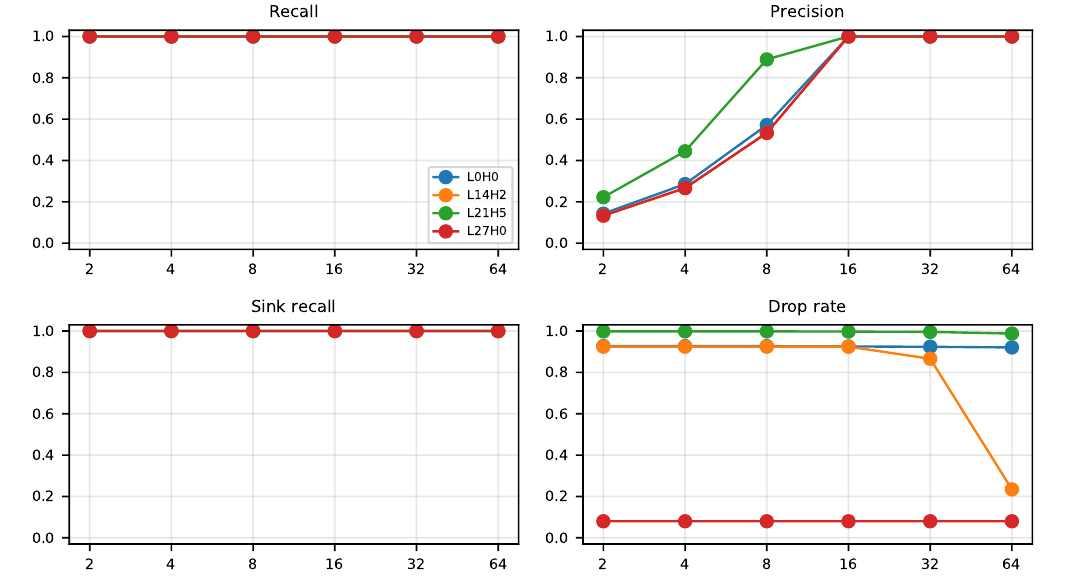}\\\footnotesize (b) DeepSeek-R1-Distill-Qwen-1.5B}
\parbox[t]{0.49\textwidth}{\centering\includegraphics[width=\linewidth]{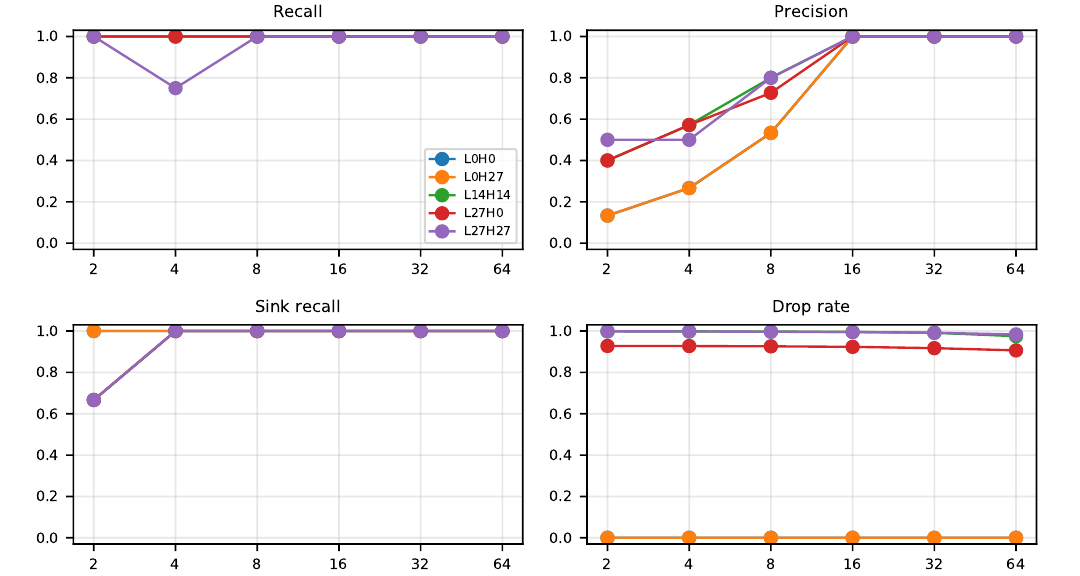}\\\footnotesize (c) Qwen2.5-7B}
\parbox[t]{0.49\textwidth}{\centering\includegraphics[width=\linewidth]{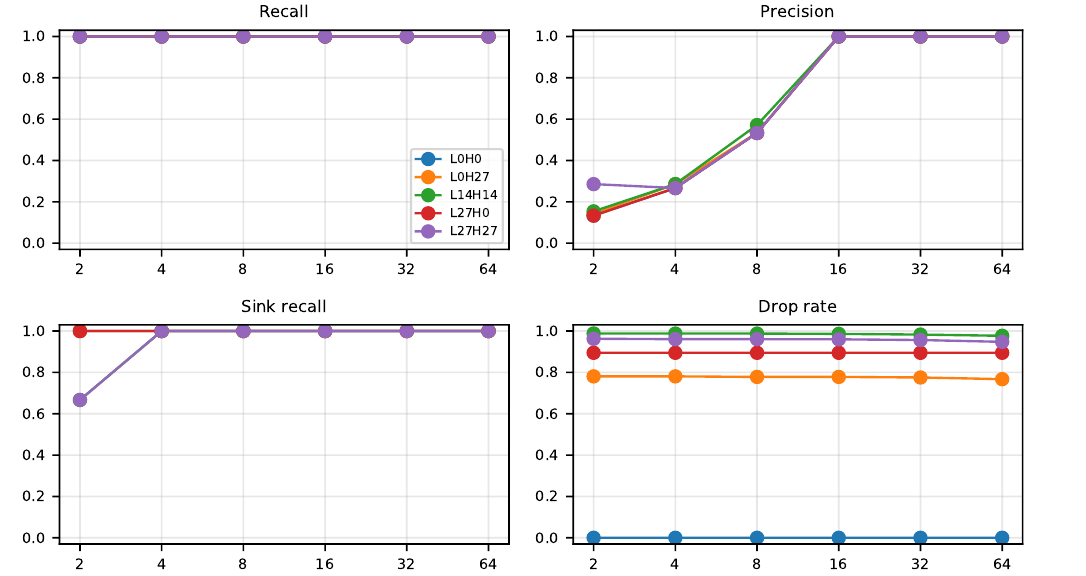}\\\footnotesize (d) DeepSeek-R1-Distill-Qwen-7B}
\caption{Empirical K-sweep trace graphs for Qwen2.5 and DeepSeek 1.5B/7B metadata streams. The four panels report recall, precision, sink recall, and drop-rate behavior versus \(K\). The finding relevant to the vector subsystem is that high maximum-score recall is obtained for practical \(K\) values while the threshold path still removes a large fraction of low-score metadata events.}
\label{fig:k_sweep_traces}
\end{figure*}

Fig.~\ref{fig:k_sweep_traces} adds the empirical trace graphs used to support the grouped-scheduling result. Across Qwen2.5 and DeepSeek traces, increasing \(K\) improves the recovered top-score set while the threshold mechanism continues to filter low-score events. These plots are especially useful for interpreting Table~\ref{tab:gqa_eval}: the system is not attempting to preserve every token position, but to retain the high-score metadata that would otherwise drive downstream KV-cache residency or candidate forwarding.

\subsection{Full DeepSeek-R1 1.5B Archive Sweep}
The full DeepSeek-R1 1.5B archive contains 336 CSV traces arranged as 28 layers and 12 heads per layer, with 8192 events per layer/head file. The evaluation script \texttt{eval/run\_deepseek\_full\_eval.py} streams the zip archive directly and applies the vector RTL identity convention: \texttt{token\_id=0} and \texttt{ptr=position}. It generates per-trace, per-layer, and per-\(K\) CSV summaries, plus graphs for recall, precision, sink retention, drop rate, and layer-by-head heatmaps. The generated artifact directory is \texttt{eval/deepseek\_full\_eval/}.

For the detected 12-head configuration, the sixteen-lane vector1024 wrapper covers one complete layer in one vector wave, with twelve active lanes and four idle lanes. Across all 28 layers, this corresponds to 229376 vector cycles for 8192 events per layer at one event per cycle per active lane. At 200 MHz, the projected metadata sweep time is 1146.88 $\mu$s, compared with 13762.56 $\mu$s for a single scalar tile processing all 336 traces sequentially. This is a scheduling projection for the existing vector wrapper, not a measured full-model inference latency.

Table~\ref{tab:deepseek_full} summarizes the aggregate full-archive sweep. Mean maximum-score recall remains high across the entire archive. The low precision values are expected because the resident table intentionally holds many more than \(K\) metadata entries; precision is therefore a resident-set purity metric rather than a direct top-\(K\) classifier score. The nonzero drop rate indicates that the threshold gate is suppressing low-score metadata traffic.

\begin{table}[t]
\centering
\caption{Full DeepSeek-R1 1.5B archive sweep over 28 layers, 12 heads/layer, and 8192 events/head using vector-mode \texttt{ptr=position} identity.}
\label{tab:deepseek_full}
\begin{tabular}{rrrrr}
\toprule
$K$ & Mean recall & Min recall & Sink recall & Drop rate \\
\midrule
4  & 0.996 & 0.250 & 0.715 & 0.405 \\
8  & 0.996 & 0.250 & 0.715 & 0.403 \\
16 & 0.996 & 0.188 & 0.711 & 0.393 \\
32 & 0.992 & 0.125 & 0.709 & 0.371 \\
64 & 0.989 & 0.531 & 0.703 & 0.340 \\
\bottomrule
\end{tabular}
\end{table}

\begin{figure}[t]
\centering
\includegraphics[width=\columnwidth]{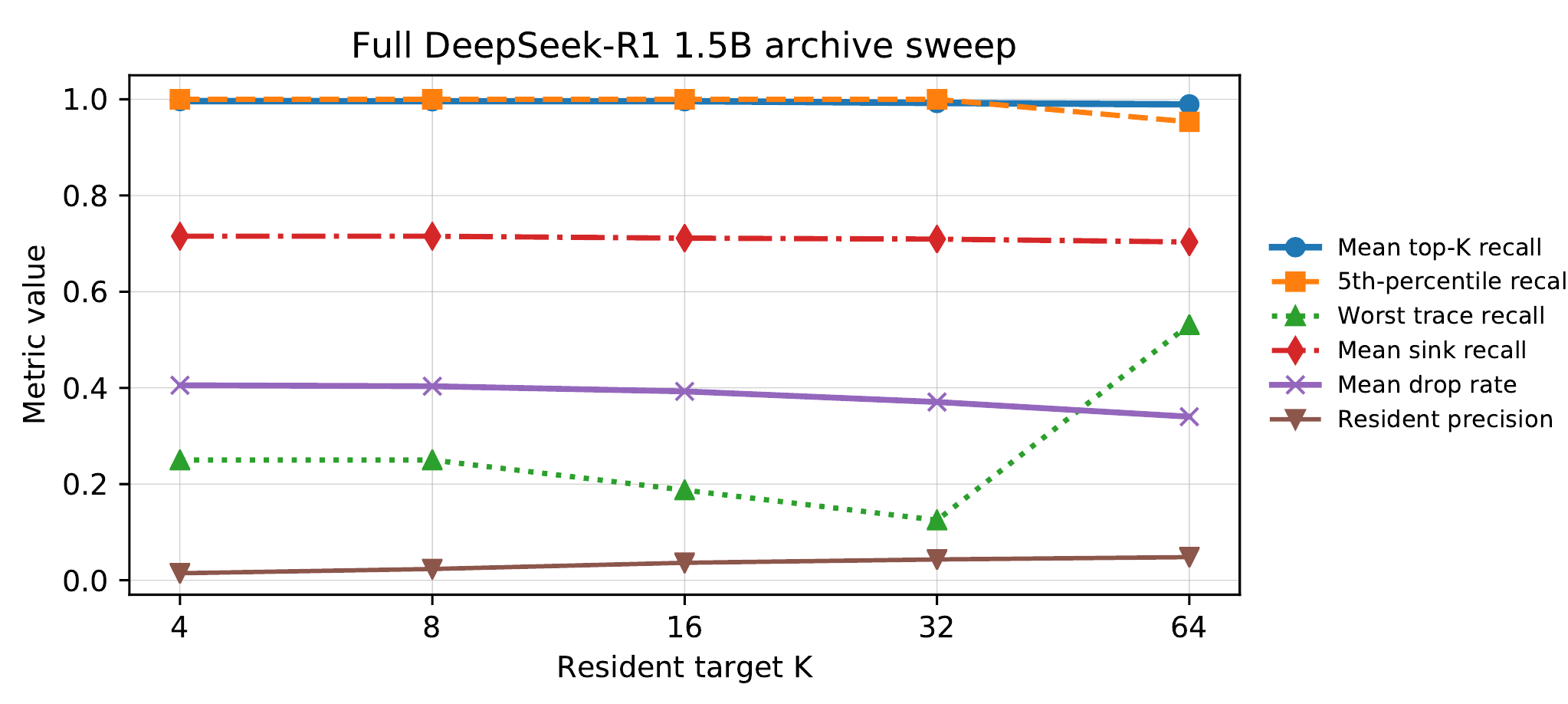}
\caption{Aggregate trends from the full DeepSeek-R1 1.5B archive sweep over 336 layer/head traces. Mean maximum-score recall remains near 0.99, while the threshold drop rate falls as \(K\) increases. The graph summarizes the CSV data in Table~\ref{tab:deepseek_full}.}
\label{fig:deepseek_full_trends}
\end{figure}

Fig.~\ref{fig:deepseek_full_trends} visualizes the full-archive aggregate trends. The main finding is that mean maximum-score recall stays near 0.99 over all reported \(K\) values, while drop rate decreases from 0.405 at \(K=4\) to 0.340 at \(K=64\). This behavior is consistent with a metadata-residency fabric: larger \(K\) values keep a broader resident set and therefore reduce filtering, but the threshold gate still blocks a substantial fraction of low-score traffic. The minimum-recall column should be interpreted conservatively. It captures worst-case layer/head traces under the current threshold/stash policy, not a model-quality failure rate; at \(K=64\), the worst observed recall improves to 0.531, while mean recall remains 0.989.

\begin{figure*}[t]
\centering
\includegraphics[width=0.49\textwidth]{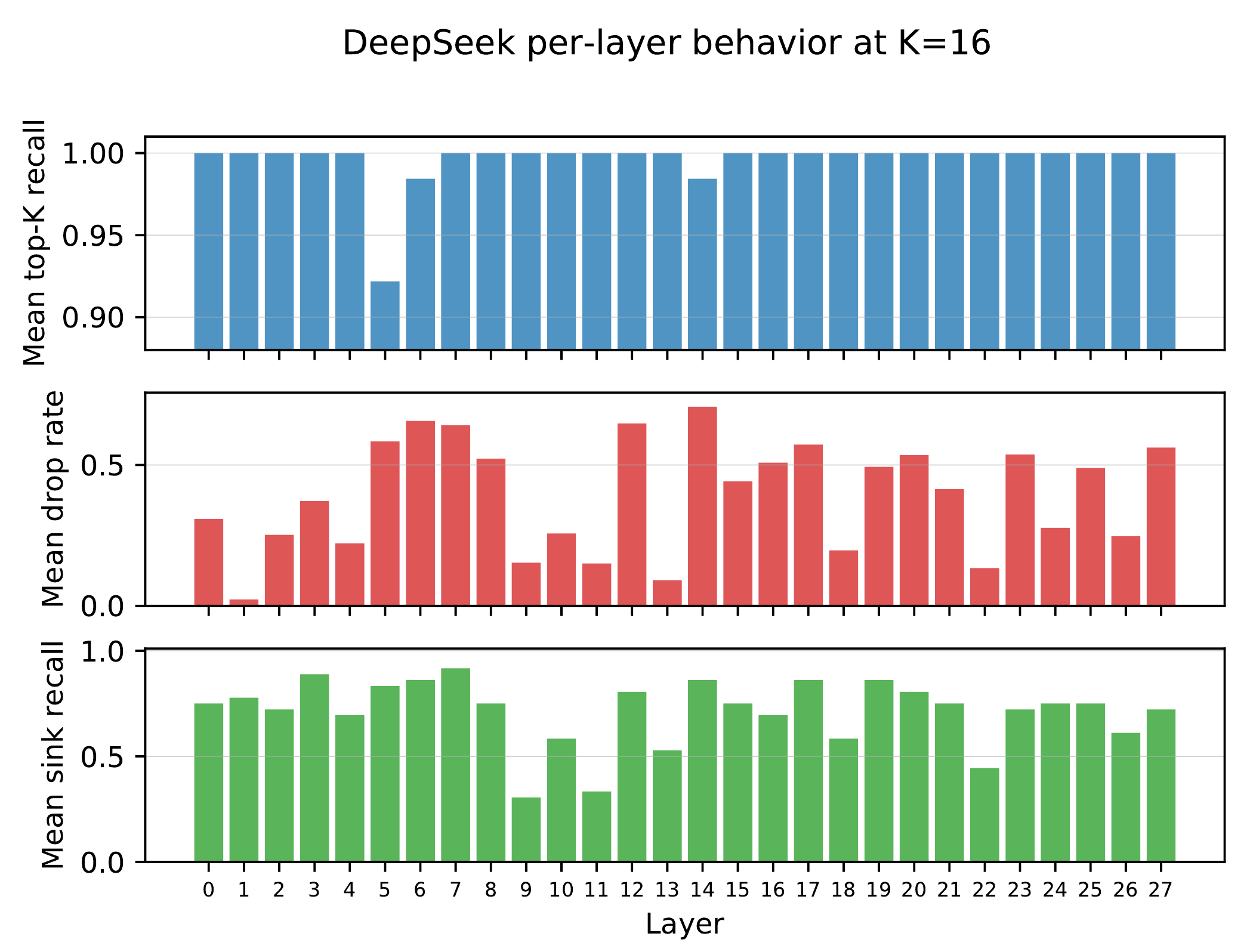}\hfill
\includegraphics[width=0.49\textwidth]{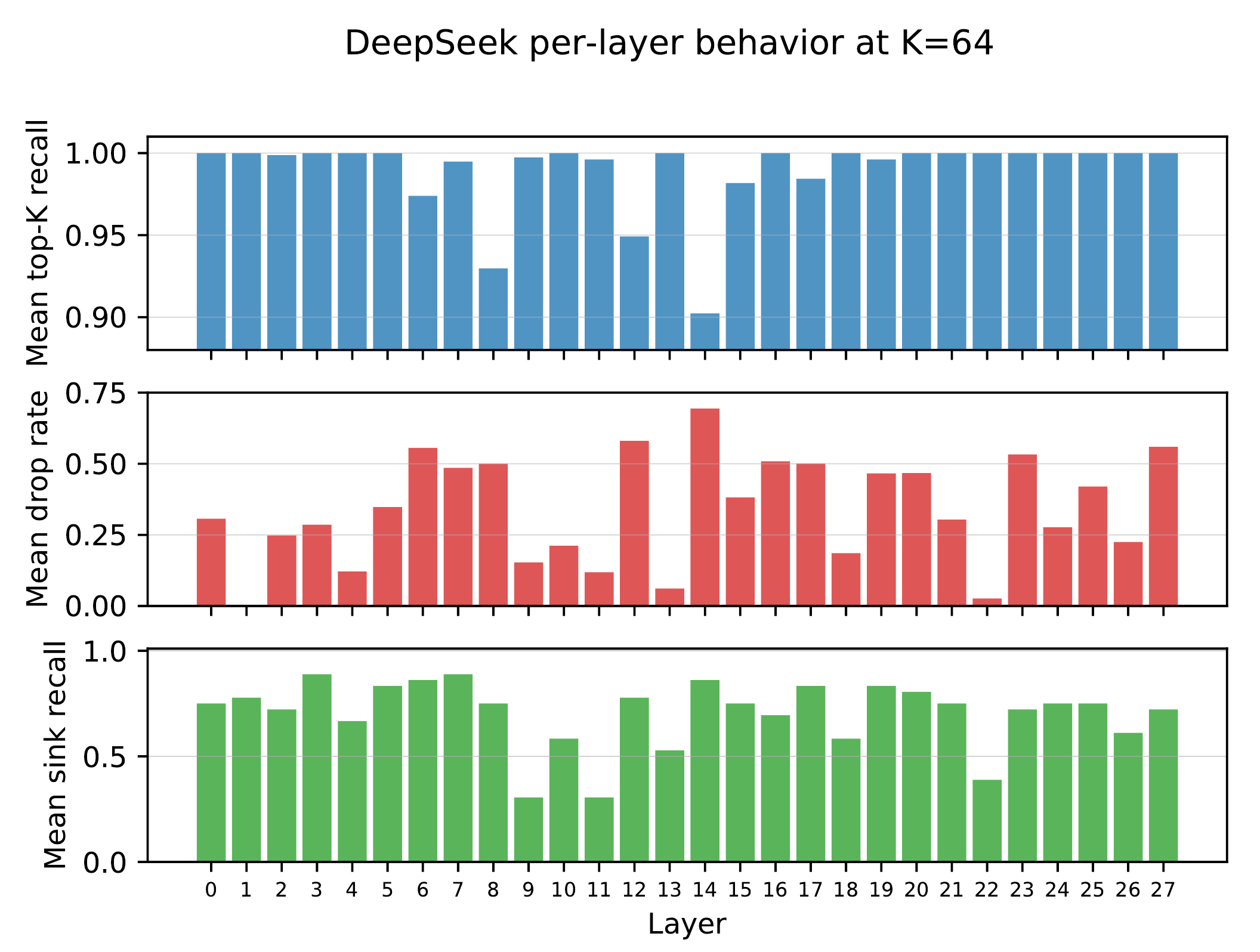}
\caption{Per-layer behavior in the full DeepSeek-R1 1.5B archive sweep at \(K=16\) and \(K=64\). At \(K=16\), layer-mean recall ranges from 0.922 to 1.000 and drop rate ranges from 0.023 to 0.707; at \(K=64\), layer-mean recall ranges from 0.902 to 1.000 and drop rate ranges from 0.001 to 0.694. These plots expose where the policy is most aggressive while preserving high average recall.}
\label{fig:deepseek_layer_bars}
\end{figure*}

\subsection{Summary of Findings}
The results support four hardware-oriented findings. First, the sixteen-lane vector1024 architecture preserves timing after replication: the OOC synthesis result meets the 5 ns target with +0.255 ns WNS, 0.000 ns TNS, and no DSP usage. Second, the lane format is well matched to metadata traffic because a 1024-bit beat carries sixteen independent 64-bit records, allowing a whole 12-head DeepSeek layer to be handled in one wave with four lanes idle. Third, for a hypothetical 128-query-head layer, the same hardware exposes the system-level value of upstream GQA aggregation: eight raw query-head waves reduce to one KV-group wave under an 8:1 mapping, provided that aggregation is performed before this block. Fourth, the full-archive trace sweep shows that the present policy maintains high mean maximum-score recall while intentionally dropping 34--41\% of metadata events across the tested \(K\) range.

These findings also define the limits of the claim. The precision numbers in Table~\ref{tab:deepseek_full} should not be compared with classifier precision; they measure how pure the resident table is relative to an oracle top-\(K\) set while the hardware intentionally keeps more than \(K\) candidates. Sink recall is computed with the current first-three-position proxy, so it is reported only as an attention-sink retention proxy rather than guaranteed BOS retention for every model head. Finally, the simplified stash is primarily a robustness mechanism for collisions and adversarial bursts; on the empirical traces used here, stash activity is small compared with threshold filtering and table admission.

\subsection{Baseline and Softmax Projection}
The scalar baseline comparison on a 50K-event Zipfian stream shows the policy tradeoff. An aggressive threshold setting retains 96.9\% maximum-score recall with 0.330 precision and 1.41\% admission, while a permissive warmup setting obtains 100\% recall with 11.31\% admission. LRU and static-threshold baselines reach full recall only by admitting every event into the evaluated path, producing much lower precision. These results support the design choice of a score-aware admission gate rather than a pure recency cache.

The softmax analysis is intentionally marked as analytical. For \(N=4096\) and \(K=64\), forwarding only \(K\) candidates reduces a downstream candidate set by 98.4375\%. At 200 MHz, the simple full-softmax-plus-top-K model is 82.24 $\mu$s, the sequential metadata scan plus candidate softmax is 21.46 $\mu$s, and an overlapped-tail estimate is 0.98 $\mu$s. These numbers are projections of a possible pre-softmax integration, not measured GPU or full-inference speedups.

\subsection{Capped-Residency Software-Model Run}
\begin{figure}[t]
\centering
\includegraphics[width=\columnwidth]{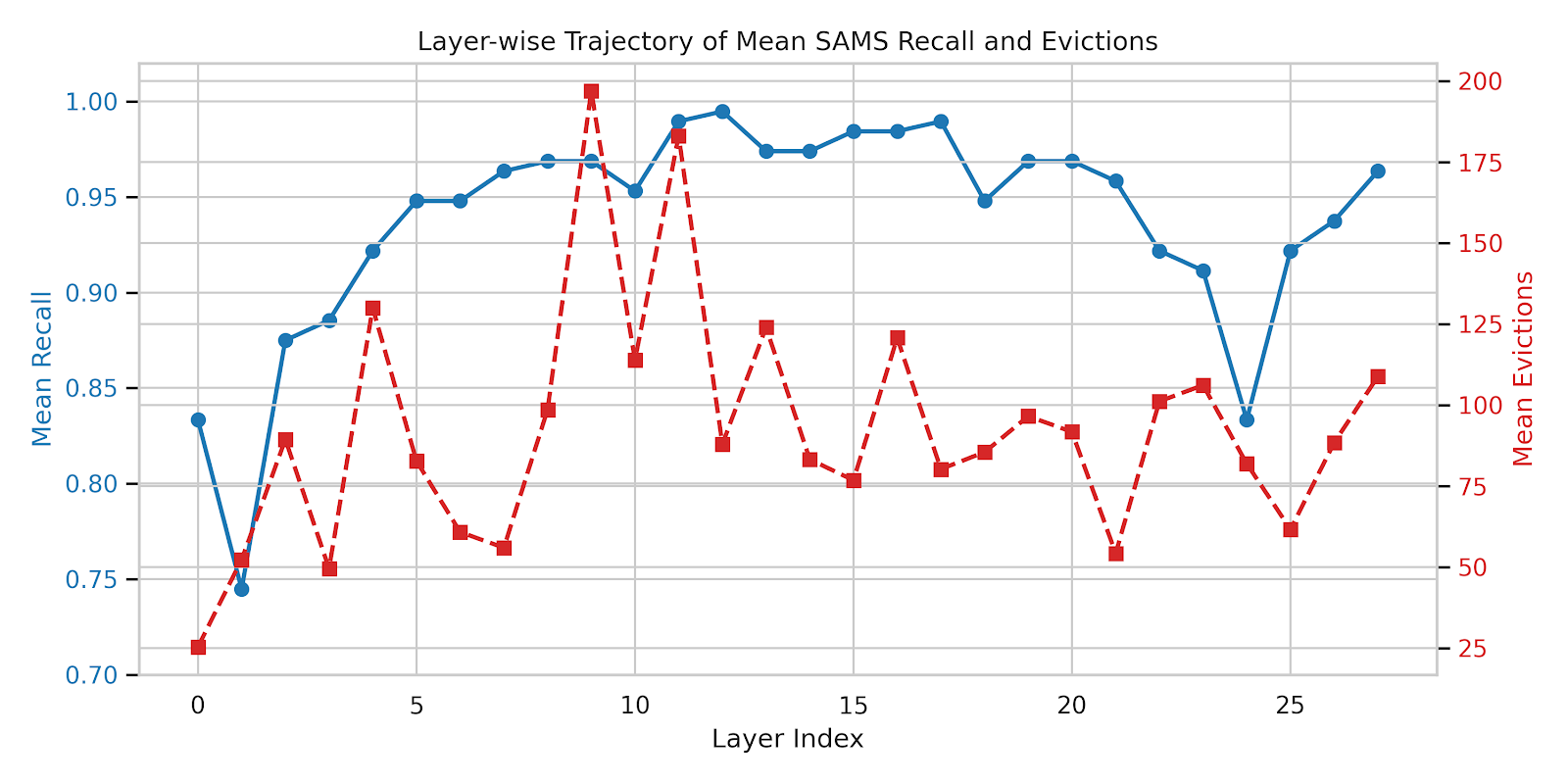}
\caption{Per-layer mean top-$K$ recall and mean eviction count in the capped-residency software-model run ($K=16$).}
\label{fig:capped}
\end{figure}

A separate software-model run covers the same 336 layer/head traces of DeepSeek-R1-Distill-Qwen-1.5B, with 8192 events per trace and 16 resident entries recorded per trace ($K=16$). Its mean admission rate is 1.32\% and its mean drop rate is 98.68\%. Mean top-$K$ recall is 93.69\%, median recall is 100\%, and the minimum over individual traces is 31.25\%. Fig.~\ref{fig:capped} shows the per-layer means. This run uses a different policy configuration from the sweep in Table~\ref{tab:deepseek_full}, whose drop rate at $K=16$ is 0.393, so the two sets of results are not directly comparable. The capped-residency configuration is not tied to the synthesis measurements in Table~\ref{tab:vector_hw}.

Comparisons with software KV-cache methods such as H$_2$O, SnapKV, Ada-KV, and Quest must distinguish algorithmic behavior from line-rate hardware datapaths; no such head-to-head comparison is reported here. Exact full attention remains the quality baseline, and downstream task evaluation is needed. The reported filtering rates count metadata events; they are not measured GPU speedups or DRAM bandwidth reductions.

\section{Discussion}
The total-system interpretation is a metadata-management result. The hardware makes deterministic local decisions at line rate and can be replicated across vector lanes. It does not replace the attention engine, softmax engine, KV-cache payload memory, or model scheduler. The current vector1024 wrapper also does not implement upstream GQA aggregation. Instead, it provides a lane-parallel metadata-residency substrate that can consume one lane per query head or one lane per KV group.

The distinction between query heads and KV groups is important. In raw query-head mode, 128 query heads need eight 1024-bit waves. In 8:1 GQA mode, sixteen KV groups fit in one 1024-bit wave, but only after the upstream attention block has produced group-level metadata. The traced distilled-Qwen models have native ratios of 6:1 (1.5B) and 7:1 (7B), so their 8:1 grouped results are scheduling approximations. Models using MLA would need a separately verified mapping.

Another limitation is storage width. The measured vector subsystem uses the existing scalar tile, which still stores 72-bit slots internally. The lane input is 64 bits, but the scalar tile reconstructs a 72-bit internal record with \(token\_id=0\). A practical future optimization is a 48-bit internal slot containing \(ptr[23:0]\), \(score[15:0]\), and \(flags[7:0]\). This optimization should reduce BRAM pressure and simplify identity handling, but it is not included in the measured resource table.

\section{Conclusion}
This paper evaluated a total vector metadata-residency subsystem for FPGA accelerator fabrics. The measured vector1024 implementation instantiates sixteen scalar metadata-residency tiles behind a 1024-bit AXI interface and meets a 200 MHz OOC synthesis target with +0.255 ns WNS, 28698 LUTs, 25955 registers, 48 RAMB36E2 blocks, and no DSPs. The functional vector smoke test passes. The GQA scheduling evaluation shows that the same sixteen lanes require eight waves for 128 raw query heads, but can cover sixteen 8:1 KV groups in one wave when upstream group aggregation is available. On available Qwen2.5 and DeepSeek trace metadata, the grouped evaluation reaches full maximum-score recall by \(K=16\) while maintaining full sink recall. The full DeepSeek-R1 1.5B archive sweep further shows that the current vector-mode SAMS model can process the detected 28-layer, 12-head trace grid in one vector wave per layer, with high mean maximum-score recall across all 336 layer/head traces.

The result is deliberately bounded: it demonstrates a practical, non-stalling metadata-residency subsystem and a credible vector/GQA deployment path. It does not claim full LLM accuracy preservation, measured GPU speedup, or full place-and-route closure of an integrated inference accelerator. The next implementation step is a true 48-bit vector-mode scalar tile and an upstream GQA aggregator that emits per-KV-group metadata directly into the vector1024 interface.

\section*{Acknowledgment}
The author thanks K. Varshasree for discussions and review of the manuscript.

\end{document}